\documentclass{article}
\usepackage{geometry}
\date{}

\usepackage{authblk}
\usepackage{lineno}
\usepackage{float}

\usepackage{amsmath,amssymb,graphicx,color}

\usepackage{bm}

\usepackage{txfonts}
\usepackage[T1]{fontenc}
\usepackage{soul}
\usepackage{braket,commath}
\usepackage{upgreek}

\usepackage{caption}
\soulregister\ref{7}
\soulregister\eqref{7}
\soulregister\cite{7}
\soulregister\onlinecite{7}

\begin{document}

\title{\textbf{\LARGE Inherent spin-orbit locking in topological bound state in the continuum lasing}}

\author[1,6,*]{Jiajun Wang}
\author[1,6]{Xinhao Wang}
\author[1]{Zhaochen Wu}
\author[1]{Xingqi Zhao}
\author[1]{Shunben Wu}
\author[1,3,4,5,*]{Lei Shi}
\author[2,*]{Yuri Kivshar}
\author[1,3,4,5,*]{Jian Zi}

\affil[1]{\raggedright State Key Laboratory of Surface Physics, Key Laboratory of Micro- and Nano-Photonic Structures (Ministry of Education) and Department of Physics, Fudan University, Shanghai 200433, China.}
\affil[2]{Nonlinear Physics Centre, Research School of Physics, The Australian National University, Canberra ACT 2601, Australia.}
\affil[3]{Institute for Nanoelectronic devices and Quantum computing, Fudan University, Shanghai 200438, China.}
\affil[4]{Collaborative Innovation Center of Advanced Microstructures, Nanjing University, Nanjing 210093, China.}
\affil[5]{Shanghai Research Center for Quantum Sciences, Shanghai 201315, China.}
\affil[6]{These authors contributed equally: Jiajun Wang, Xinhao Wang.}
\affil[*]{e-mail: jiajunwang@fudan.edu.cn; lshi@fudan.edu.cn; yuri.kivshar@anu.edu.au; jzi@fudan.edu.cn}

\maketitle


\large
\section*{Abstract}

Bound states in the continuum (BICs) are exotic optical topological singularities that defy the typical radiation within the continuum of radiative modes and carry topological polarization vortices in momentum space. Enabling ultrahigh quality factors, BICs have been applied in realizing lasing and Bose-Einstein condensation via micro-/nano- photonic structures, and their momentum-space vortex topologies have been exploited in passive systems, revealing novel spin-orbit photonic effects. However, as representative topological properties, the spin-orbit-related phenemona of BICs in active systems have not yet been explored. Here, we demonstrate the inherent spin-orbit locking in topological BIC lasing. Utilizing photonic crystal (PhC) slabs with square (C$_{4v}$) and triangular (C$_{6v}$) lattices, we achieve distinct spin-orbit locking combinations in topological BIC lasing of $+1$ and $-2$ topological charges. These BIC lasing profiles manifest as vortex and high-order anti-vortex polarization configurations, directly tied to the topological properties of BICs. Our experimental results directly reveal the spin-orbit locking phenomena through momentum-space spin-dependent self-interference patterns and real-space spin separations of the lasing emissions. This study not only highlights the inherent spin-orbit-locking behaviours of topological BIC lasing but also opens new possibilities for dynamically switchable orbital angular momentum (OAM) lasing by controlling photonic spin, presenting significant potential for advancements in topological  photonic source applications.


\section*{Introduction}


Bound states in the continuum (BICs), initially proposed in quantum mechanics, have since established themselves as pivotal phenomena in modern photonics due to their rich physical implications~\cite{hsu2016bound,kang2023applications,wang2024optical}. With infinite quality factors, BICs stand as topological optical singularities that remian perfectly confined within the continuum of radiative modes~\cite{marinica2008bound,hsu2013observation}. Especially in periodic structures like photonic crystal (PhC) slabs, BICs are found to carry complex topological polarization vortices in momentum space~\cite{zhen2014topological,zhang2018observation,doeleman2018experimental}. 
These properties of BICs have been utilized to enhance versatile applications in various areas such as sensors~\cite{tittl2018imaging,yesilkoy2019ultrasensitive}, lasers~\cite{kodigala2017lasing,ha2018directional,huang2020ultrafast,hwang2021ultralow,sang2022topological}, Bose-Einstein condensation~\cite{ardizzone2022polariton,gianfrate2024reconfigurable}, light-matter interaction~\cite{maggiolini2023strongly,weber2023intrinsic,cotrufo2024passive}, and spin-orbit interaction of light~\cite{wang2020generating,wang2022spin}. 
Especially, the momentum-space vectiorial topology of BICs, out of the well-known topological band theory for scalar fields, gives rise to new perspectives of topological photonics in the far field. 

For optical fields in free space, the spin angular momentum and orbital angular momentum (OAM) of light are two crucial indices of light, being closely related to the circular polarizations ($\sigma_{\pm}$) and the winding number ($l$) of the spiral phase~\cite{allen1999iv,andrews2012angular,shen2019optical}. Exploring the relation and interaction between the spin angular momentum and OAM of light has attracted great attention in photonics. In vectorial optical fields, the spin and orbit are closely connected with each other, underlying many unique phenomena in spin-orbit photonics like spin-to-vortex conversion and photonic spin Hall effect~\cite{bliokh2015spin,song2015unveiling,ling2017recent,chen2018vectorial}. From this perspective, novel topological spin-orbit phenomena are enabled and expected in the momentum-space vectorial polarization vortex configurations of BICs. On the one hand, in passive systems, the spin-orbit interations of light have been explored via resonance-induced responses under circularly polarized incidence. The spin-to-vortex conversion has been proposed and demonstrated via polarization vortices around BICs, giving rise to a new nonlocal method for optical vortex generation~\cite{wang2020generating}. Furthermore, the spin Hall effect of light are proposed via BICs, in which the in-plane-oblique spin-dependent beam shifts are realized~\cite{wang2022spin}. On the other hand, in active systems, the radiations based on BICs will directly inherit the properties of BICs, which are initially explored in BIC-modulated emission such as lasing and Bose-Einstein condensation~\cite{kodigala2017lasing,ha2018directional,huang2020ultrafast,hwang2021ultralow,sang2022topological,ardizzone2022polariton,gianfrate2024reconfigurable}. However, to date, the spin-orbit radiation phenomena of BICs have not been explored.

In this work, we demonstrate the inherent spin-orbit locking in topological BIC lasing. The spin-orbit-locking lasing behaviours are inherent from the topological polarization configurations of BICs. By taking advantage of various momentum-space topological configurations of BICs enabled by point-group symmetries, we can realize different spin-orbit locking combinations in BIC lasing. To experimentally demonstrate the spin-orbit-locking properties of BIC lasing, we used photonic crystal (PhC) slabs of square lattice (C$_{4v}$) and triangular lattice (C$_{6v}$) to obtain different types of symmetry-protected BICs. Both vortex and high-order anti-vortex polarization configurations were realized in BIC microlasers with topological charges of $+1$ and $-2$. By the momentum-space spin-dependent self-interference patterns and real-space spin separations of lasing profiles, the inherent spin-orbit locking in BIC lasing was directly identified. Our results revealed new spin-orbit-locking phenomena in BIC lasing, which can be utilized to realize switchable OAM lasing by control photonic spin.

\section*{Results}

\subsection*{Polarization vortices of BICs and the inherent spin-orbit locking}

\begin{figure}[H]
 \centering
  \includegraphics[scale=1]{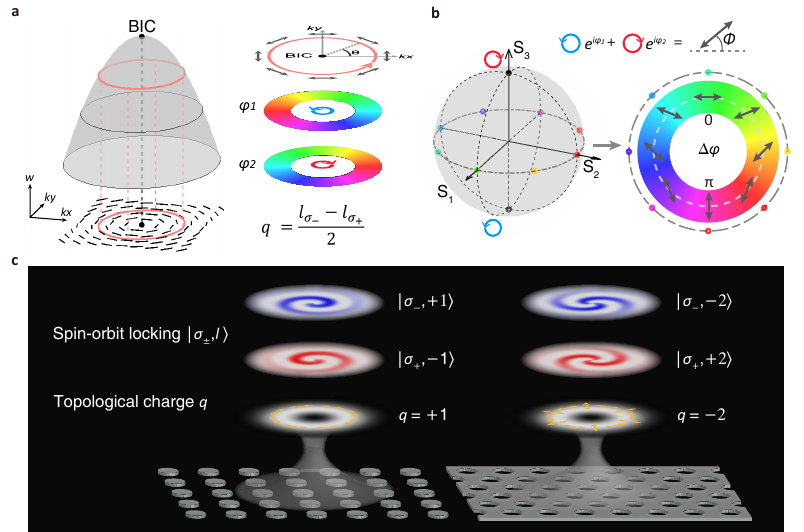}
  \caption{\textbf{$\textbar$ Concept of the spin-orbit locking in BICs.} \textbf{a}, The schematic of a photonic band and topological polarization vortex in momentum space. The BIC is at the centre of the band, surrounding which the eigen polarizations of the optical modes form vortex configuration in the projected momentum space. The topological charge $q$ of the BIC is defined by the total winding angle of the polarization orentation $\phi$ in a close loop along the anti-clockwise direction (such as the the magenta circle). This topological charge underlies inherent spin-orbit locking of BICs, corresponding to the spin-dependent phase winding numbers $l_{\sigma_\pm}$. In the loop surrouding the BIC, two spin components ($\sigma_\pm$) of polarization states carry phase vortices of opposite signs. \textbf{b}, The Poincar\'e sphere representation for spin-dependent phase vortices. The linearly polarized state on the equator of the Poincar\'e sphere can be decomposed by its two poles (circular polarizations). Winding along the equator, the polarization orentation $\phi$ of the linearly polarized state rotates continuously, corresponding to the continuous evolution of $\Delta\varphi$. \textbf{c}, The schematic view of the proposed spin-orbit locking $|\sigma_\pm, l\rangle$ in topological BIC lasing. The topological vectorial lasing of $+1$ topological charge is realzied in the PhC slab of square lattice, in whcih the spin-orbit locking is $|\sigma_\pm, \mp1\rangle$. The topological vectorial lasing of $-2$ topological charge is realzied in the PhC slab of triangular lattice, in whcih the spin-orbit locking is $|\sigma_\pm, \pm2\rangle$.}
  \label{Fig:1}
\end{figure}

Fig. \ref{Fig:1}\textbf{a} schematically depicts a photonic band structure of the PhC slab, whose center is a BIC. This BIC carries $+1$ topological charge, embodying as the momentum-space polarization vortex by projecting eigen-polarizations of optical modes in the photonic band onto the momentum ($k_x-k_y$) plane. Importantly, besides this exhibited vectorial polarization vortex, there is underlying spin-orbit related properties inherent in the topological configuration of BICs. For example, surrounding this BIC along the magenta circle in momentum space, the orentation of the polarization state rotates anti-clockwise as the azimuth angle $\theta$ increases. The $+1$ topological charge of the BIC corresponds to a total accumulated winding angle of $+2\pi$. Furthermore, when considering two spin components of these polarization states ($\sigma_\pm$), we can see that opposite spiral phase windings exist along the magenta circle in momentum space. In other words, the phase distribution of two spin components form phase vortices with opposite phase winding numbers (orbit, $l$) around the BIC. Each spin is locked with a specific OAM $l$. In the case shown in Fig. \ref{Fig:1}\textbf{a}, the left-handed circularly polarization ($\sigma_-$) carrys a positive phase optical vortex ($l = 1$), and the right-handed circularly polarization ($\sigma_+$) carrys a positive phase optical vortex ($l = -1$).

The origin of this spin-orbit locking can be understood using the Poincar\'e sphere representation, as is shown in Fig. \ref{Fig:1}\textbf{b}. We consider linear polarization state $|E(\phi)\rangle$ for an example, which can be decomposed by two poles ($|\sigma_\pm\rangle$) of the Poincar\'e sphere as:
\begin{equation}\label{eq:1}
\begin{aligned}
&|E(\phi)\rangle = e^{i\varphi_1}|\sigma_-\rangle + e^{i\varphi_2}|\sigma_+\rangle, \\
&\phi = \frac{\varphi_1-\varphi_2}{2} = \frac{\Delta\varphi}{2}.
\end{aligned}
\end{equation}
Here, $\phi$ is defined polarization orentation angle of the linear polarization state, and $\varphi_1, \varphi_2$ is the phase of the spin basis. The polarization orentation angle $\phi$ is equal to the phase difference $\Delta\varphi$ of two spin components. As is exhibited in the right panel of Fig. \ref{Fig:1}\textbf{b}, winding along the equator of the Poincar\'e sphere, $\Delta\phi$ vary continuously with $\phi$. Similarly, we can project polarization vortices carried by BICs onto the Poincar\'e sphere, then the topological charge $q$ of the BIC can be written as:
\begin{equation}\label{eq:2}
q = \frac{1}{2\pi}\oint_{C} \frac{\partial \phi}{\partial \theta} \mathrm{d}\theta = \frac{1}{4\pi}\oint_{C}(\frac{\partial \varphi_1}{\partial \theta}-\frac{\partial \varphi_2}{\partial \theta}) \mathrm{d}\theta = \frac{l_{\sigma_-}-l_{\sigma_+}}{2}.
\end{equation}
Here, C is a loop around the BIC in momentum space (such as the magenta circle in Fig. \ref{Fig:1}\textbf{a}). And $l_{\sigma_\pm}$ is the accumulated phase winding number of the $\sigma_\pm$ component. We can see the spin-orbit locking $|\sigma_\pm,l\rangle$ properties are inherent in the topological charges of BICs. For the mostly discussed symmetry-protected BICs, we consider the existence of one mirror symmetry, then $l_{\sigma_-}$ is equal to $-l_{\sigma_+}$. We can simplify Eq. \ref{eq:2} as: 
\begin{equation}\label{eq:3}
q = l_{\sigma_-} = -l_{\sigma_+}.
\end{equation}
From this perspective, various spin-orbit locking $|\sigma_\pm,l\rangle$ exist in BICs with different topological charges, which can be realized by manipulating symmetries. Based on this result, we can also class the spin-orbit locking by topological charges of BICs, and vise versa.,

Here, we further propose that, as the significant application of BICs, BIC lasing also have the inherent spin-orbit locking properties. Fig. \ref{Fig:1}\textbf{c} exhibits schematics of two examples of BIC lasing with distinct spin-orbit locking behaviors. The first one utilizes the BIC of $+1$ topological charge in a $C_{4v}$ PhC slab. The total lasing profile is a vortex-polarization-configuration vectorial beam, and the inherent spin-orbit locking relations are $|\sigma_\pm, \mp1\rangle$. It means the $\sigma_-$ lasing component is a vortex beam with $+2\pi$ spiral phase winding and the $\sigma_+$ lasing component is a vortex beam with $-2\pi$ spiral phase winding. The second one utilizes the BIC of $-2$ topological charge in a $C_{6v}$ PhC slab. The total lasing profile is an anti-vortex-polarization-configuration vectorial beam, and the inherent spin-orbit locking relations are $|\sigma_\pm, \pm2\rangle$.

\subsection*{Design of topological BIC lasing}

To realize the spin-orbit-locking phenomena in BIC lasing, we designed two types of PhC slabs with different symmetries, specifically square lattice (C$_{4v}$) and triangular lattice (C$_{6v}$). The PhC structures are composed of silicon nitride (Si$_3$N$_4$, refractive index $\sim$ 2) in an environment of optical silica (refractive index $\sim$ 1.45). By utilizing finite-element method simulation, we obtained the photonic band structures and polarization distributions in momentum space to ensure that designed PhC structures align with the experimental condition. It should be noted that two symmetries are chosen to exhibit types of BICs with different topological charges. The relations between symmetries and topological charges see Supplementary Information Section 1.

\begin{figure}[htpb]
 \centering
  \includegraphics[scale=1]{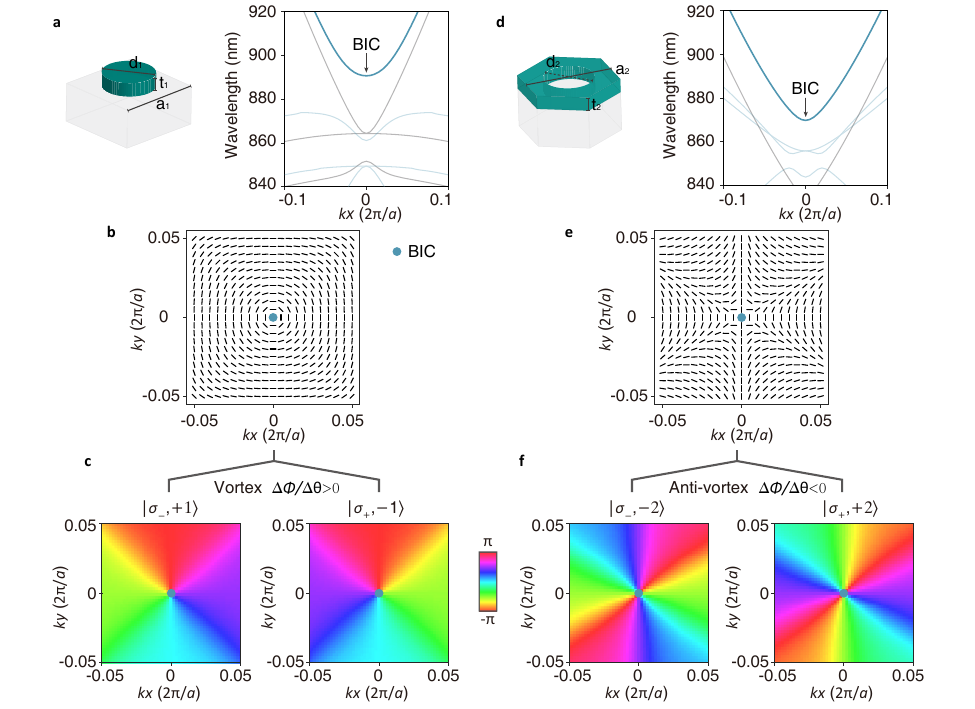}
  \caption{$\textbar$ \textbf{Simulated topological polarization vortex and spin-dependent phase vortex of BICs in C$_{4v}$ and  C$_{6v}$ PhC slabs.} \textbf{a}, The unit cell of the designed C$_{4v}$ PhC slab and the simulated photonic bands. The arrow marks the central BIC on the focused TE-like band. \textbf{b}, The momentum-space polarization vortex on the cyan-color-marked photonic band in \textbf{a}, in which the topological charge of the BIC is $+1$. \textbf{c}, The spin-dependent phase distributions in momentum space. Surrounding the BIC of $+1$ topological charge, the spiral phase winding number $l$ of the $\sigma_-$ component is $+1$, and that of the $\sigma_+$ componentis $-1$. \textbf{d}, The unit cell of the designed C$_{6v}$ PhC slab and the simulated photonic bands. The arrow marks the central BIC on the focused TE-like band. \textbf{e}, The momentum-space polarization vortex on the cyan-color-marked photonic band in \textbf{d}, in which the topological charge of the BIC is $-2$. \textbf{f}, The spin-dependent phase distributions in momentum space. Surrounding the BIC of $-2$ topological charge,the spiral phase winding number $l$ of the $\sigma_-$ component is $-2$, and that of the $\sigma_+$ component is $+2$. }
  \label{Fig:2}
\end{figure}

Fig. \ref{Fig:2}\textbf{a} shows the schematic of the C$_{4v}$ PhC slab. The unit cell consist of a circular Si$_3$N$_4$ pillar with lattice constant $a_1$ of 560 nm, pillar diameter $d_1$ of 420 nm, and slab thickness $t_1$ of 240 nm. The corresponding photonic band structure, depicted in the right panel of Fig. \ref{Fig:2}\textbf{a}, demonstrates a symmetry-protected BIC at the $\Gamma$ point with a topological charge of $+1$. Fig. \ref{Fig:2}\textbf{b} shows the momentum-space polarization distribution around the BIC. The polarization states are nearly linearly polarized and form vortex configuration around the BIC. Then, we calculated momentum-space phase distributions of two spin components, as are shown in Fig. \ref{Fig:2}\textbf{c}. The results show $\sigma_-$ components carries a positive phase vortex ($l = +1$) and $\sigma_+$ component carries a negative phase vortex ($l = -1$), agreeing well with the Eq. \ref{eq:3}.
Similarly, Fig. \ref{Fig:2}\textbf{d} illustrates the schematic of the C$_{6v}$ PhC slab, composed of a triangular array of etched circular holes in a Si$_3$N$_4$ film. The lattice constant $a_2$ is 634 nm, hole diameter $d_2$ is 320 nm, and slab thickness $t_2$ is 120 nm. The right panel of Fig. \ref{Fig:2}\textbf{d} shows the corresponding photonic band structures, in which the arrow marks the required symmetry-protected BIC at the $\Gamma$ point with a topological charge of $-2$. Fig. \ref{Fig:2}\textbf{e} shows the anti-vortex polarization configuration around the BIC. The calculated momentum-space phase distributions in Fig. \ref{Fig:2}\textbf{f} show that $\sigma_-$ component carries a positive phase vortex ($l = -2$) and $\sigma_+$ component carries a negative phase vortex ($l = +2$).
In these designed PhC structures, two types of BICs are obtained to include both vortex and anti-vortex polarization configurations and two distinct spin-orbit-locking behaviours.

\subsection*{The topological polarization configurations BIC lasing}

\begin{figure}[H]
\centering
\includegraphics[scale=1]{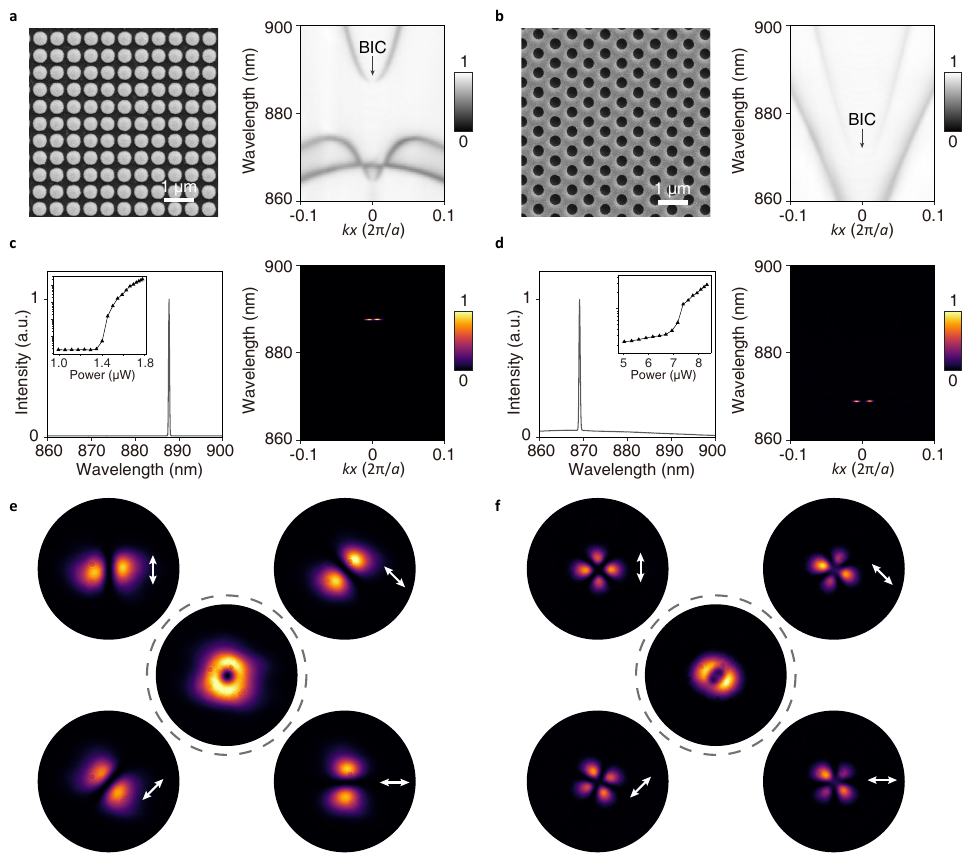}
\caption{$\textbar$ \textbf{The experimental topological BIC lasing and the vectorial polarization vortex characterizations.} \textbf{a}, \textbf{b}, Scanning electron microscopy images of the fabricated PhC slabs and corresponding measured momentum-resolved transmittance spectra. \textbf{a}, C$_{4v}$ PhC slab. \textbf{b}, C$_{6v}$ PhC slab. \textbf{c}, \textbf{d},
 Measured above-threshold integrated lasing spectra and momentum-resolved lasing spectra via PhC slabs of \textbf{a} and \textbf{b}. The insets shows input-output integrated photoluminescence curves. a.u., arbitrary units. \textbf{e}, \textbf{f}, Measured polarization-resolved lasing images in momentum space via PhC slabs of \textbf{a} and \textbf{b}. The double arrows show directions of the analyzed polarizations. The images inside the dashed circles exhibit lasing images in the experimental setup without the polarizer.  }
  \label{Fig:3}
\end{figure}

To experimentally implement the proposed spin-orbit locking in BIC lasing, we then fabricated the designed two types of PhC slabs. Figs. \ref{Fig:3}\textbf{a} and \textbf{b} show the scanning electron microscopy images and corresponding measured momentum-resolved transmittance spectra of fabricated C$_{4v}$ and C$_{6v}$ samples respectively. The black arrows mark the designed two types of BICs in Fig. \ref{Fig:2}.

To realize BIC lasing, we chose IR-140 dye molecules as the gain, which are dissolved in dimethyl sulfoxide to match the refractive index of the optical silica. The lasing measurements were performed by a built momentum-space spectropy system. A femtosecond laser (800 nm, $\sim$100 fs pulse width at a repetition rate of 1 kHz) was applied to optically pump the designed samples with liquid gain, with a laser spot of $\sim 50$ $\upmu$m focused by a $\times 4$ microscope objective. Figs. \ref{Fig:3}\textbf{c} and \textbf{d} exhibit the measured above-threshold integrated lasing spectra, input-output integrated photoluminescence curves (insets) and momentum-resolved lasing spectra corresponding to samples shown in Figs. \ref{Fig:3}\textbf{a} and \textbf{b}. Obvious lasing features were observed including characteristic threshold behaviours and sharp peaks. Additionally, for the BIC lasing, the banned emissions at the normal direction were also observed due to the nonradiative nature of BICs.

Figs. \ref{Fig:3}\textbf{e} and \textbf{f} show the momentum-space BIC lasing profiles of $+1$ and $-2$ topological charges (Fig. \ref{Fig:2}\textbf{b} and \textbf{e}). The total lasing profiles, the images inside the dashed circle, show the doughnut shape. By adding a polarizer before the camera, polarization properties of lasing profiles could be directly measured. The momentum-space topological polarization configurations of the BIC lasing were examined by taking polarization-resolved images. For the lasing profile via BIC of $+1$ topological charge, there are two nodes for certain linear polarization. And as polarization direction rotates anticlockwise, the nodes rotate along the same dirction, agreeing well with the vortex polarization configuration in Fig. \ref{Fig:2}\textbf{b}. For the lasing profile via BIC of $-2$ topological charge, there are four nodes for certain linear polarization. And as polarization direction rotates anticlockwise, the nodes roate along the opposite dirction, agreeing well with the anti-vortex polarization configuration in Fig. \ref{Fig:2}\textbf{e}. More polarization-analyzed results are shown in Supplementary  Fig. S7 and S8 to exhibit continuous variations.

\subsection*{Observation of the inherent spin-orbit locking}

To directly observe the spin-orbit locking in these BIC lasing, we performed self-interference experiments and gave corresponding numerical results for comparison. In the experiments, the BIC lasing profiles were split into two parts and finally combined. The interferences happened on the camera plane due to slightly off-separate centres of two profiles. The self-interference patterns were measured in three types of setups, in which the off-centre lasing profiles are the total, $\sigma_-$ and $\sigma_+$ components. The detailed experimental setups see Supplementary Information Section 3.

\begin{figure}[H]
 \centering
  \includegraphics[scale=1]{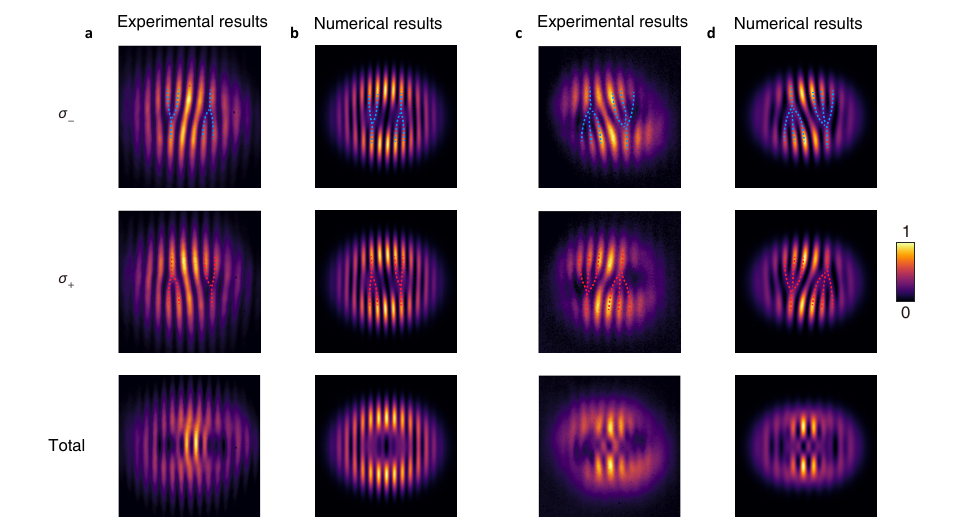}
  \caption{$\textbar$ \textbf{The direct observation of the spin-orbit locking in topological BIC lasing.} \textbf{a}-\textbf{d}, Measured self-interference patterns of the lasing profiles. \textbf{a}, \textbf{b}, The experimental and numerical results for the lasing profiles in Fig. 3\textbf{e}, whose topological charge is $+1$. \textbf{c}, \textbf{d}, The experimental and numerical results for the lasing profiles in Fig. 3\textbf{f}, whose topological charge is $-2$. The images of the first and second rows correspond to self-interference results of the $\sigma_-$ and $\sigma_+$ components, respectively. The fork-shaped fringes which are related to vortex phases are marked with dashed lines. The images of the last row correspond to self-interference results of the total profiles, in which no fork-shaped patterns were observed.}
  \label{Fig:4}
\end{figure}

In self-interference patterns of the $\sigma_\mp$ components (the first and second rows in  Fig. \ref{Fig:4}), fork-shaped interference fringes were observed. Dashed lines marked the fringe splitting. For the lasing profiles via the BIC of $+1$ topological charge (Figs. \ref{Fig:4}\textbf{a} and \textbf{b}), one fringe is split into two in the centres of lasing profiles, indicating the absolute OAM $|l|$ is equal to 1. Especially, when changing from $\sigma_-$ to $\sigma_+$, we can see the direction of fork-shaped fringes reverses, corresponding to reversed signs of the OAM $l$. In comparsion with numerical results, we can see the $\sigma_-$ lasing profile carries an OAM of $+1$  and the $\sigma_+$ lasing profile carries an OAM of $-1$. For the lasing profiles via the BIC of $-2$ topological charge (Figs. \ref{Fig:4}\textbf{c} and \textbf{d}), one fringe is split into three in the centres of lasing profiles, indicating the absolute OAM $|l|$ is equal to 2. The spin-orbit locking was also observed where the $\sigma_-$ lasing profile carries an OAM of $-2$ and the $\sigma_+$ lasing profile carries an OAM of $+2$. Notably, for the same spin components ($\sigma_-$ or $\sigma_+$) of lasing profiles via BICs of $+1$ and $-2$ topological charges, the directions of fork-shaped fringes are opposite to each other, demonstrating different spin-orbit-locking behaviours governed by topological charges of BICs (Eq. \ref{eq:3}). 

In contrast, for the self-interference patterns of the total lasing profiles, there are not such obvious fork-shaped interference patterns, meaning the absence of spiral phases. The interference patterns are nearly mirror-symmetric due to the existed mirror symmetry of the polarization vortex configurations, agreeing well with the numerical results. Based on previous discussions in theoretical parts, we can know two spin-orbit coupled components in conventional BICs are degenerate, hence the total OAM are zero. From this perspective, we also experimentally demonstrated the zero total OAM in conventional BIC lasing. This also inspires the future development of BIC lasing with non-zero OAM by introducing additional symmetry breaking to change the distribution of two spin components or lift the spin degeneracy~\cite{qin2023arbitrarily,zhao2024spin}.

\begin{figure}[H]
 \centering
  \includegraphics[scale=1]{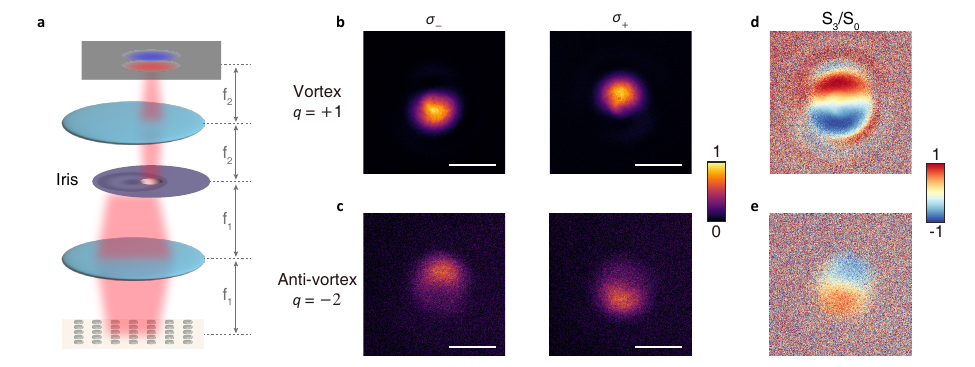}
  \caption{$\textbar$ \textbf{The spin separations of lasing profiles in real space.} \textbf{a}, The schematic view of the experimental setup. An iris is put on the Fourier plane to filter part of the lasing profiles in momentum space, and the images in the real space are captured by the camera. \textbf{b}, \textbf{c}, The experimental observation of spin separations in real space for the filtered lasing profiles in Fig. 3\textbf{e} and Fig. 3\textbf{f}. The filtered regions in momentum space are as shown in the schematic of \textbf{a}. Scale bars: 100 $\upmu$m. \textbf{d}, \textbf{e}, The normalized third Stokes parameters in real space calculated from \textbf{b} and \textbf{c}.}
  \label{Fig:5}
\end{figure}

Moreover, besides the spin-related self-interference patterns in momentum space, the spin-orbit locking can also manifest as spin separations in real space. In the discussions of spin-orbit locking based on the topologcial polarization configurations, we have known that two spin components carry opposite OAM in momentum space, i.e., opposite tangential phase gradients.  Based on the fact that momentum space and real space are a pair of reciprocal spaces, such spin-dependent momentum-space phase gradients correspond to spin-dependent separations of real-space position $\langle \bm{R}_{\sigma_\pm} \rangle$ as~\cite{wang2022spin,ling2017recent,wang2021shifting}:  
\begin{equation}\label{eq:4}
\langle \bm{R}_{\sigma_\pm} \rangle = - \langle \frac{\partial \varphi_{\sigma_\pm}(\bm{k}_{||})}{\partial \bm{k}_{||}} \rangle.
\end{equation}
Here, $\langle \bm{R}_{\sigma_\pm} \rangle$ is the lasing profile in real space, $\bm{k}_{||}$ is the projected in-plane momentum and $\varphi_{\sigma_\pm}(\bm{k}_{||})$ is the phase of $\sigma_\pm$-component lasing profile at certain $\bm{k}_{||}$. To observe the expected spin separations, we built a real-space measurement system and Fig. \ref{Fig:5}\textbf{a} exhibits the schematic of the experimental setup. An iris is put on the Fourier plane relative to the sample, and only one part of doughnut-shape lasing profiles was filtered to pass through. The spin-dependent lasing profiles $I_{\sigma_\pm}$ were captured by a camera put on the real-image plane of the sample, as are shown in Figs. \ref{Fig:5}\textbf{b} and \textbf{c}. To clearly exhibit the spin separations, we further calculated the distribution of the normalized third Stokes parameter by $\rm{\frac{S_3}{S_0}} = \frac{I_{\sigma_+}-I_{\sigma_-}}{I_{\sigma_+}+I_{\sigma_-}}$, as are shown in Figs. \ref{Fig:5}\textbf{d} and \textbf{e}. The clear spin separations were observed. There are opposite spin separations for BIC lasing of $+1$ topological charge (Figs. \ref{Fig:5}\textbf{b} and \textbf{d}) and $-2$ topological charge (Figs. \ref{Fig:5}\textbf{c} and \textbf{e}), showing good accordance with the spin-orbit-locking behaviours of two types of BICs. These observed spin separations reveal properties of the BIC lasing in the near field. The BIC lasing has spin-dependent near-field distributions of emission profiles, and manifests as spin-orbit locking in far field.   

The unique spin-orbit locking properties in BIC lasing are important for both BICs and lasing research areas. In previous lasing studies via BICs, the high Q factors are firstly explored to realize lasing~\cite{kodigala2017lasing,ha2018directional,yu2021ultra,ren2022low}, then the topological polarization configurations of BICs are applied to realize vectorial polarization vortex lasing~\cite{huang2020ultrafast,hwang2021ultralow,sang2022topological,wu2020room,mohamed2022controlling}. Here, beyond the topological polarization configurations, the inherent spin-orbit locking in BICs is further discovered and utilized to realize various spin-orbit-locking lasing. The spin-orbit-locking lasing, due to the potential in light-field manipulation and information processing, has attracted much research attention in the last decade~\cite{shen2019optical,forbes2024orbital}. Lasing with spin-orbit-locking properties has been reported in composite laser cavities integrated with wave plates or metasurfaces~\cite{naidoo2016controlled,sroor2020high,piccardo2022vortex}, designed benzene-like photonic molecules~\cite{carlon2019optically}, topological photonic crystals~\cite{yang2020spin} and non-Hermitian-coupling-modulated microring cavities~\cite{zhang2020tunable,zhang2022spin}. Among them, miniaturization is one development trend for the microlasers, and the BIC lasers via PhC slabs can possess advantages in developing compact microlasers. In addition, previous spin-orbit-locking lasers usually have structural centers, resulting in accurate alignment requirements in both fabrications and applications. In contrast, the PhC slabs supporting BICs are periodic structures, enabling the advantages of no real-space alignment. BIC lasing offers a new nonlocal method to generate spin-orbit-locking lasing, and the topological charges of BICs also offer a new degree of freedom to manipulte both the spin-orbit realtions and the orders of the orbital angular momentum. Moreover, the spin-orbit locking properties enable the BIC lasers to realize switchable optical vortices by controlling spins, offering promising applications of vortex beams.

\section*{Conclusion}

In this work, we have explored the inherent spin-orbit locking in topological BIC lasing using PhC slabs with different symmetries. Our findings demonstrate that the spin-orbit locking behavior of BICs can be harnessed to achieve switchable OAM lasing. The experimental results confirmed the presence of spin-dependent phase vortices and real-space spin separations in the lasing profiles, directly tied to the topological properties of the BICs. These results not only deepen our understanding of the interplay between spin and orbital angular momentum in photonic systems but also pave the way for novel applications in spin-controlled photonic devices and topological photonics. Future work will focus on exploring the potential of these phenomena in more complex photonic structures and their integration into practical devices.

\bibliography{Ref}

\begin{thebibliography}{10}
\expandafter\ifx\csname url\endcsname\relax
  \def\url#1{\texttt{#1}}\fi
\expandafter\ifx\csname urlprefix\endcsname\relax\def\urlprefix{URL }\fi
\providecommand{\bibinfo}[2]{#2}
\providecommand{\eprint}[2][]{\url{#2}}

\bibitem{hsu2016bound}
\bibinfo{author}{Hsu, C.~W.}, \bibinfo{author}{Zhen, B.},
  \bibinfo{author}{Stone, A.~D.}, \bibinfo{author}{Joannopoulos, J.~D.} \&
  \bibinfo{author}{Solja{\v{c}}i{\'c}, M.}
\newblock \bibinfo{title}{Bound states in the continuum}.
\newblock \emph{\bibinfo{journal}{Nature Reviews Materials}}
  \textbf{\bibinfo{volume}{1}}, \bibinfo{pages}{1--13} (\bibinfo{year}{2016}).

\bibitem{kang2023applications}
\bibinfo{author}{Kang, M.}, \bibinfo{author}{Liu, T.}, \bibinfo{author}{Chan,
  C.} \& \bibinfo{author}{Xiao, M.}
\newblock \bibinfo{title}{Applications of bound states in the continuum in
  photonics}.
\newblock \emph{\bibinfo{journal}{Nature Reviews Physics}}
  \textbf{\bibinfo{volume}{5}}, \bibinfo{pages}{659--678}
  (\bibinfo{year}{2023}).

\bibitem{wang2024optical}
\bibinfo{author}{Wang, J.} \emph{et~al.}
\newblock \bibinfo{title}{Optical bound states in the continuum in periodic
  structures: mechanisms, effects, and applications}.
\newblock \emph{\bibinfo{journal}{Photonics Insights}}
  \textbf{\bibinfo{volume}{3}}, \bibinfo{pages}{R01--R01}
  (\bibinfo{year}{2024}).

\bibitem{marinica2008bound}
\bibinfo{author}{Marinica, D.}, \bibinfo{author}{Borisov, A.} \&
  \bibinfo{author}{Shabanov, S.}
\newblock \bibinfo{title}{Bound states in the continuum in photonics}.
\newblock \emph{\bibinfo{journal}{Physical review letters}}
  \textbf{\bibinfo{volume}{100}}, \bibinfo{pages}{183902}
  (\bibinfo{year}{2008}).

\bibitem{hsu2013observation}
\bibinfo{author}{Hsu, C.~W.} \emph{et~al.}
\newblock \bibinfo{title}{Observation of trapped light within the radiation
  continuum}.
\newblock \emph{\bibinfo{journal}{Nature}} \textbf{\bibinfo{volume}{499}},
  \bibinfo{pages}{188--191} (\bibinfo{year}{2013}).

\bibitem{zhen2014topological}
\bibinfo{author}{Zhen, B.}, \bibinfo{author}{Hsu, C.~W.}, \bibinfo{author}{Lu,
  L.}, \bibinfo{author}{Stone, A.~D.} \& \bibinfo{author}{Solja{\v{c}}i{\'c},
  M.}
\newblock \bibinfo{title}{Topological nature of optical bound states in the
  continuum}.
\newblock \emph{\bibinfo{journal}{Physical review letters}}
  \textbf{\bibinfo{volume}{113}}, \bibinfo{pages}{257401}
  (\bibinfo{year}{2014}).

\bibitem{zhang2018observation}
\bibinfo{author}{Zhang, Y.} \emph{et~al.}
\newblock \bibinfo{title}{Observation of polarization vortices in momentum
  space}.
\newblock \emph{\bibinfo{journal}{Physical review letters}}
  \textbf{\bibinfo{volume}{120}}, \bibinfo{pages}{186103}
  (\bibinfo{year}{2018}).

\bibitem{doeleman2018experimental}
\bibinfo{author}{Doeleman, H.~M.}, \bibinfo{author}{Monticone, F.},
  \bibinfo{author}{den Hollander, W.}, \bibinfo{author}{Al{\`u}, A.} \&
  \bibinfo{author}{Koenderink, A.~F.}
\newblock \bibinfo{title}{Experimental observation of a polarization vortex at
  an optical bound state in the continuum}.
\newblock \emph{\bibinfo{journal}{Nature Photonics}}
  \textbf{\bibinfo{volume}{12}}, \bibinfo{pages}{397--401}
  (\bibinfo{year}{2018}).

\bibitem{tittl2018imaging}
\bibinfo{author}{Tittl, A.} \emph{et~al.}
\newblock \bibinfo{title}{Imaging-based molecular barcoding with pixelated
  dielectric metasurfaces}.
\newblock \emph{\bibinfo{journal}{Science}} \textbf{\bibinfo{volume}{360}},
  \bibinfo{pages}{1105--1109} (\bibinfo{year}{2018}).

\bibitem{yesilkoy2019ultrasensitive}
\bibinfo{author}{Yesilkoy, F.} \emph{et~al.}
\newblock \bibinfo{title}{Ultrasensitive hyperspectral imaging and biodetection
  enabled by dielectric metasurfaces}.
\newblock \emph{\bibinfo{journal}{Nature Photonics}}
  \textbf{\bibinfo{volume}{13}}, \bibinfo{pages}{390--396}
  (\bibinfo{year}{2019}).

\bibitem{kodigala2017lasing}
\bibinfo{author}{Kodigala, A.} \emph{et~al.}
\newblock \bibinfo{title}{Lasing action from photonic bound states in
  continuum}.
\newblock \emph{\bibinfo{journal}{Nature}} \textbf{\bibinfo{volume}{541}},
  \bibinfo{pages}{196--199} (\bibinfo{year}{2017}).

\bibitem{ha2018directional}
\bibinfo{author}{Ha, S.~T.} \emph{et~al.}
\newblock \bibinfo{title}{Directional lasing in resonant semiconductor
  nanoantenna arrays}.
\newblock \emph{\bibinfo{journal}{Nature nanotechnology}}
  \textbf{\bibinfo{volume}{13}}, \bibinfo{pages}{1042--1047}
  (\bibinfo{year}{2018}).

\bibitem{huang2020ultrafast}
\bibinfo{author}{Huang, C.} \emph{et~al.}
\newblock \bibinfo{title}{Ultrafast control of vortex microlasers}.
\newblock \emph{\bibinfo{journal}{Science}} \textbf{\bibinfo{volume}{367}},
  \bibinfo{pages}{1018--1021} (\bibinfo{year}{2020}).

\bibitem{hwang2021ultralow}
\bibinfo{author}{Hwang, M.-S.} \emph{et~al.}
\newblock \bibinfo{title}{Ultralow-threshold laser using super-bound states in
  the continuum}.
\newblock \emph{\bibinfo{journal}{Nature Communications}}
  \textbf{\bibinfo{volume}{12}}, \bibinfo{pages}{4135} (\bibinfo{year}{2021}).

\bibitem{sang2022topological}
\bibinfo{author}{Sang, Y.-G.} \emph{et~al.}
\newblock \bibinfo{title}{Topological polarization singular lasing with highly
  efficient radiation channel}.
\newblock \emph{\bibinfo{journal}{Nature communications}}
  \textbf{\bibinfo{volume}{13}}, \bibinfo{pages}{6485} (\bibinfo{year}{2022}).

\bibitem{ardizzone2022polariton}
\bibinfo{author}{Ardizzone, V.} \emph{et~al.}
\newblock \bibinfo{title}{Polariton bose--einstein condensate from a bound
  state in the continuum}.
\newblock \emph{\bibinfo{journal}{Nature}} \textbf{\bibinfo{volume}{605}},
  \bibinfo{pages}{447--452} (\bibinfo{year}{2022}).

\bibitem{gianfrate2024reconfigurable}
\bibinfo{author}{Gianfrate, A.} \emph{et~al.}
\newblock \bibinfo{title}{Reconfigurable quantum fluid molecules of bound
  states in the continuum}.
\newblock \emph{\bibinfo{journal}{Nature Physics}}
  \textbf{\bibinfo{volume}{20}}, \bibinfo{pages}{61--67}
  (\bibinfo{year}{2024}).

\bibitem{maggiolini2023strongly}
\bibinfo{author}{Maggiolini, E.} \emph{et~al.}
\newblock \bibinfo{title}{Strongly enhanced light--matter coupling of monolayer
  ws2 from a bound state in the continuum}.
\newblock \emph{\bibinfo{journal}{Nature Materials}}
  \textbf{\bibinfo{volume}{22}}, \bibinfo{pages}{964--969}
  (\bibinfo{year}{2023}).

\bibitem{weber2023intrinsic}
\bibinfo{author}{Weber, T.} \emph{et~al.}
\newblock \bibinfo{title}{Intrinsic strong light-matter coupling with
  self-hybridized bound states in the continuum in van der waals metasurfaces}.
\newblock \emph{\bibinfo{journal}{Nature Materials}}
  \textbf{\bibinfo{volume}{22}}, \bibinfo{pages}{970--976}
  (\bibinfo{year}{2023}).

\bibitem{cotrufo2024passive}
\bibinfo{author}{Cotrufo, M.}, \bibinfo{author}{Cordaro, A.},
  \bibinfo{author}{Sounas, D.~L.}, \bibinfo{author}{Polman, A.} \&
  \bibinfo{author}{Al{\`u}, A.}
\newblock \bibinfo{title}{Passive bias-free non-reciprocal metasurfaces based
  on thermally nonlinear quasi-bound states in the continuum}.
\newblock \emph{\bibinfo{journal}{Nature Photonics}}
  \textbf{\bibinfo{volume}{18}}, \bibinfo{pages}{81--90}
  (\bibinfo{year}{2024}).

\bibitem{wang2020generating}
\bibinfo{author}{Wang, B.} \emph{et~al.}
\newblock \bibinfo{title}{Generating optical vortex beams by momentum-space
  polarization vortices centred at bound states in the continuum}.
\newblock \emph{\bibinfo{journal}{Nature Photonics}}
  \textbf{\bibinfo{volume}{14}}, \bibinfo{pages}{623--628}
  (\bibinfo{year}{2020}).

\bibitem{wang2022spin}
\bibinfo{author}{Wang, J.}, \bibinfo{author}{Shi, L.} \& \bibinfo{author}{Zi,
  J.}
\newblock \bibinfo{title}{Spin hall effect of light via momentum-space
  topological vortices around bound states in the continuum}.
\newblock \emph{\bibinfo{journal}{Physical Review Letters}}
  \textbf{\bibinfo{volume}{129}}, \bibinfo{pages}{236101}
  (\bibinfo{year}{2022}).

\bibitem{allen1999iv}
\bibinfo{author}{Allen, L.}, \bibinfo{author}{Padgett, M.} \&
  \bibinfo{author}{Babiker, M.}
\newblock \bibinfo{title}{Iv the orbital angular momentum of light}.
\newblock In \emph{\bibinfo{booktitle}{Progress in optics}},
  vol.~\bibinfo{volume}{39}, \bibinfo{pages}{291--372}
  (\bibinfo{publisher}{Elsevier}, \bibinfo{year}{1999}).

\bibitem{andrews2012angular}
\bibinfo{author}{Andrews, D.~L.} \& \bibinfo{author}{Babiker, M.}
\newblock \emph{\bibinfo{title}{The angular momentum of light}}
  (\bibinfo{publisher}{Cambridge University Press}, \bibinfo{year}{2012}).

\bibitem{shen2019optical}
\bibinfo{author}{Shen, Y.} \emph{et~al.}
\newblock \bibinfo{title}{Optical vortices 30 years on: Oam manipulation from
  topological charge to multiple singularities}.
\newblock \emph{\bibinfo{journal}{Light: Science \& Applications}}
  \textbf{\bibinfo{volume}{8}}, \bibinfo{pages}{90} (\bibinfo{year}{2019}).

\bibitem{bliokh2015spin}
\bibinfo{author}{Bliokh, K.~Y.}, \bibinfo{author}{Rodr{\'\i}guez-Fortu{\~n}o,
  F.~J.}, \bibinfo{author}{Nori, F.} \& \bibinfo{author}{Zayats, A.~V.}
\newblock \bibinfo{title}{Spin--orbit interactions of light}.
\newblock \emph{\bibinfo{journal}{Nature Photonics}}
  \textbf{\bibinfo{volume}{9}}, \bibinfo{pages}{796--808}
  (\bibinfo{year}{2015}).

\bibitem{song2015unveiling}
\bibinfo{author}{Song, D.} \emph{et~al.}
\newblock \bibinfo{title}{Unveiling pseudospin and angular momentum in photonic
  graphene}.
\newblock \emph{\bibinfo{journal}{Nature communications}}
  \textbf{\bibinfo{volume}{6}}, \bibinfo{pages}{6272} (\bibinfo{year}{2015}).

\bibitem{ling2017recent}
\bibinfo{author}{Ling, X.} \emph{et~al.}
\newblock \bibinfo{title}{Recent advances in the spin hall effect of light}.
\newblock \emph{\bibinfo{journal}{Reports on Progress in Physics}}
  \textbf{\bibinfo{volume}{80}}, \bibinfo{pages}{066401}
  (\bibinfo{year}{2017}).

\bibitem{chen2018vectorial}
\bibinfo{author}{Chen, J.}, \bibinfo{author}{Wan, C.} \& \bibinfo{author}{Zhan,
  Q.}
\newblock \bibinfo{title}{Vectorial optical fields: recent advances and future
  prospects}.
\newblock \emph{\bibinfo{journal}{Science Bulletin}}
  \textbf{\bibinfo{volume}{63}}, \bibinfo{pages}{54--74}
  (\bibinfo{year}{2018}).

\bibitem{qin2023arbitrarily}
\bibinfo{author}{Qin, H.} \emph{et~al.}
\newblock \bibinfo{title}{Arbitrarily polarized bound states in the continuum
  with twisted photonic crystal slabs}.
\newblock \emph{\bibinfo{journal}{Light: Science \& Applications}}
  \textbf{\bibinfo{volume}{12}}, \bibinfo{pages}{66} (\bibinfo{year}{2023}).

\bibitem{zhao2024spin}
\bibinfo{author}{Zhao, X.} \emph{et~al.}
\newblock \bibinfo{title}{Spin-orbit-locking chiral bound states in the
  continuum}.
\newblock \emph{\bibinfo{journal}{Physical Review Letters}}
  \textbf{\bibinfo{volume}{133}}, \bibinfo{pages}{036201}
  (\bibinfo{year}{2024}).

\bibitem{wang2021shifting}
\bibinfo{author}{Wang, J.} \emph{et~al.}
\newblock \bibinfo{title}{Shifting beams at normal incidence via controlling
  momentum-space geometric phases}.
\newblock \emph{\bibinfo{journal}{Nature Communications}}
  \textbf{\bibinfo{volume}{12}}, \bibinfo{pages}{6046} (\bibinfo{year}{2021}).

\bibitem{yu2021ultra}
\bibinfo{author}{Yu, Y.} \emph{et~al.}
\newblock \bibinfo{title}{Ultra-coherent fano laser based on a bound state in
  the continuum}.
\newblock \emph{\bibinfo{journal}{Nature Photonics}}
  \textbf{\bibinfo{volume}{15}}, \bibinfo{pages}{758--764}
  (\bibinfo{year}{2021}).

\bibitem{ren2022low}
\bibinfo{author}{Ren, Y.} \emph{et~al.}
\newblock \bibinfo{title}{Low-threshold nanolasers based on miniaturized bound
  states in the continuum}.
\newblock \emph{\bibinfo{journal}{Science Advances}}
  \textbf{\bibinfo{volume}{8}}, \bibinfo{pages}{eade8817}
  (\bibinfo{year}{2022}).

\bibitem{wu2020room}
\bibinfo{author}{Wu, M.} \emph{et~al.}
\newblock \bibinfo{title}{Room-temperature lasing in colloidal nanoplatelets
  via mie-resonant bound states in the continuum}.
\newblock \emph{\bibinfo{journal}{Nano Letters}} \textbf{\bibinfo{volume}{20}},
  \bibinfo{pages}{6005--6011} (\bibinfo{year}{2020}).

\bibitem{mohamed2022controlling}
\bibinfo{author}{Mohamed, S.} \emph{et~al.}
\newblock \bibinfo{title}{Controlling topology and polarization state of lasing
  photonic bound states in continuum}.
\newblock \emph{\bibinfo{journal}{Laser \& Photonics Reviews}}
  \textbf{\bibinfo{volume}{16}}, \bibinfo{pages}{2100574}
  (\bibinfo{year}{2022}).

\bibitem{forbes2024orbital}
\bibinfo{author}{Forbes, A.}, \bibinfo{author}{Mkhumbuza, L.} \&
  \bibinfo{author}{Feng, L.}
\newblock \bibinfo{title}{Orbital angular momentum lasers}.
\newblock \emph{\bibinfo{journal}{Nature Reviews Physics}}
  \textbf{\bibinfo{volume}{6}}, \bibinfo{pages}{352–364}
  (\bibinfo{year}{2024}).

\bibitem{naidoo2016controlled}
\bibinfo{author}{Naidoo, D.} \emph{et~al.}
\newblock \bibinfo{title}{Controlled generation of higher-order poincar{\'e}
  sphere beams from a laser}.
\newblock \emph{\bibinfo{journal}{Nature Photonics}}
  \textbf{\bibinfo{volume}{10}}, \bibinfo{pages}{327--332}
  (\bibinfo{year}{2016}).

\bibitem{sroor2020high}
\bibinfo{author}{Sroor, H.} \emph{et~al.}
\newblock \bibinfo{title}{High-purity orbital angular momentum states from a
  visible metasurface laser}.
\newblock \emph{\bibinfo{journal}{Nature Photonics}}
  \textbf{\bibinfo{volume}{14}}, \bibinfo{pages}{498--503}
  (\bibinfo{year}{2020}).

\bibitem{piccardo2022vortex}
\bibinfo{author}{Piccardo, M.} \emph{et~al.}
\newblock \bibinfo{title}{Vortex laser arrays with topological charge control
  and self-healing of defects}.
\newblock \emph{\bibinfo{journal}{Nature Photonics}}
  \textbf{\bibinfo{volume}{16}}, \bibinfo{pages}{359--365}
  (\bibinfo{year}{2022}).

\bibitem{carlon2019optically}
\bibinfo{author}{Carlon~Zambon, N.} \emph{et~al.}
\newblock \bibinfo{title}{Optically controlling the emission chirality of
  microlasers}.
\newblock \emph{\bibinfo{journal}{Nature Photonics}}
  \textbf{\bibinfo{volume}{13}}, \bibinfo{pages}{283--288}
  (\bibinfo{year}{2019}).

\bibitem{yang2020spin}
\bibinfo{author}{Yang, Z.-Q.}, \bibinfo{author}{Shao, Z.-K.},
  \bibinfo{author}{Chen, H.-Z.}, \bibinfo{author}{Mao, X.-R.} \&
  \bibinfo{author}{Ma, R.-M.}
\newblock \bibinfo{title}{Spin-momentum-locked edge mode for topological vortex
  lasing}.
\newblock \emph{\bibinfo{journal}{Physical review letters}}
  \textbf{\bibinfo{volume}{125}}, \bibinfo{pages}{013903}
  (\bibinfo{year}{2020}).

\bibitem{zhang2020tunable}
\bibinfo{author}{Zhang, Z.} \emph{et~al.}
\newblock \bibinfo{title}{Tunable topological charge vortex microlaser}.
\newblock \emph{\bibinfo{journal}{Science}} \textbf{\bibinfo{volume}{368}},
  \bibinfo{pages}{760--763} (\bibinfo{year}{2020}).

\bibitem{zhang2022spin}
\bibinfo{author}{Zhang, Z.} \emph{et~al.}
\newblock \bibinfo{title}{Spin--orbit microlaser emitting in a four-dimensional
  hilbert space}.
\newblock \emph{\bibinfo{journal}{Nature}} \textbf{\bibinfo{volume}{612}},
  \bibinfo{pages}{246--251} (\bibinfo{year}{2022}).

\end{thebibliography}
\bibliographystyle{naturemag}

\section*{Methods}
\noindent \textbf{Simulations.} The photonic band simulations and the polarization/phase analysis were performed by using a finite-element method (COMSOL Multiphysics). The periodic boundary conditions were applied in the x and y directions and the second-order scattering boundary condition was applied in the z direction.

\noindent \textbf{Sample fabrication.} The samples were fabricated on the Si$_3$N$_4$ film. The Si$_3$N$_4$ film was grown on the optical silica substrate by PECVD (Oxford PlasmaPro system 100). The thickness of the Si$_3$N$_4$ film could be controlled by growth time. A CSAR 62 positive electron beam resist layer followed by a conductive polymer layer (AR-PC 5092.02) was spin-coated on the Si$_3$N$_4$ film. By using electron beam lithography (JEOL JBX-8100FS), the patterned mask was made on the electron beam resist layer. For the C$_{6v}$ PhC slab, the patterned electron beam resist was applied as the etching mask. For the C$_{4v}$ PhC slab, a 15-nm-thick Cr layer was further deposited on the patterned electron beam resist and lifted off by acetone to generate Cr cylinder arrays as the etching mask. The patterns were then transferred from mask to the Si$_3$N$_4$ film by reactive ion etching (RIE, Trion T2) using a mixture of CHF$_3$ and O$_2$. By removing the mask, the PhC slabs were finally prepared. Please see the Supplementary Information Section 2 for the detailed fabrication processes of two types of PhC slabs in this work.

\noindent \textbf{Optical measurements.} The optical mesurements were performed by a built Fourier-optics-based spectroscopy and imaging system. The measurement system has three working modes: a spectrometer mode, an imaging mode and a self-interference mode. The momentum-resolved and wavelength-resolved spectra could be measured by the spectrometer mode. For the transmittance spectra measurement, a broadband white light was applied as the light source~\cite{zhang2018observation}. For the lasing measurement, a femtosecond laser (800 nm, $\sim$100 fs pulse width at a repetition rate of 1 kHz) was applied as the pumping light and a 850-nm long-pass filter was applied to filter the pumping light. For the imaging mode, a camera was put on the Fourier plane relative to the sample plane. The momentum-space lasing profiles were obtained by the imaging mode. A linear polarizer was applied to realize the polarization-analyzed imaging. The self-interference mode was based on a built Michelson interferometer module. The lasing profile was split into two and recombined by the same beam splitter. By modulating the mirrors, off-centre combined lasing profiles could form interference patterns and be recorded by the camera~\cite{zhang2020tunable}. Please see the Supplementary Information Section 3 for the schemtic view and detailed descriptions of the optical measurement system.

The measurments of real-space spin separations were based on the same system with an additional lens on the imaging mode. The schematic view was shown in Fig. \ref{Fig:5}(a). A linear polarizer and quarter-wave plate were applied to realize spin-dependent imaging. 

\section*{Data availability}
The data that support the plots within this paper and other findings of this study are available from the corresponding authors upon reasonable request.

\section*{Acknowledgements}
This work was supported by National Key R\&D Program of China (No. 2023YFA1406900 and No. 2022YFA1404800); National Natural Science Foundation of China (No. 12234007, No. 12321161645, and No. 12221004); Major Program of National Natural Science Foundation of China (Grants No. T2394480, No. T2394481); Science and Technology Commission of Shanghai Municipality (22142200400, 21DZ1101500, 2019SHZDZX01 and 23DZ2260100). J.W. was further supported by China National Postdoctoral Program for Innovative Talents (BX20230079) and China Postdoctoral Science Foundation (2023M740721).

\section*{Author contributions}
J.W., L.S., Y.K. and J.Z. conceived the basic idea and designed the experiments. J.W., X.W. and X.Z. gave the theoretical explanation and performed numerical simulations. X.W. fabricated samples. X.W. and J.W. constructed the measurement systems. X.W., J.W., Z.W and S.W. performed the optical measurements. J.W., X.W., L.S., Y.K. and J.Z. analysed the data. J.W. wrote the manuscript, and all authors took part in the discussions and revisions and approved the final copy of the manuscript.

\section*{Competing interests}
The authors declare no competing interests.

\end{document}